\documentclass{aastex62}
\usepackage{amsmath}
\usepackage{wasysym}
\usepackage{wrapfig}
\usepackage{graphicx}
\usepackage{hyperref}
\usepackage{float}
\usepackage{ulem}
\usepackage{xcolor}
\usepackage{epstopdf}
\usepackage[caption=false]{subfig}

\begin{document}
\title{Investigating the Temperature Distribution of Diatomic Carbon in Comets using the Swan Bands}
\shorttitle{C$_2$ temperatures in comets 122P and 153P}
\author[0000-0003-3707-5746]{Tyler Nelson}
\affiliation{Astronomy Department, The University of Texas at Austin, 2515 Speedway, Austin, TX, 78712, USA}
\author[0000-0003-4828-7787]{Anita L. Cochran}
\affiliation{McDonald Observatory, The University of Texas at Austin, 2515 Speedway, Austin, TX, 78712, USA}

\shortauthors{Nelson, Cochran}
\correspondingauthor{Tyler Nelson}
\email{tyler.nelson@utexas.edu}
\submitjournal{AJ}
\received{08/27/2019}
\revised{10/08/2019}
\accepted{10/08/2019}
\begin{abstract}
We present high spectral-resolution observations of comets 122P/de~Vico and 153P/Ikeya-Zhang obtained with the Tull Coud\'{e} spectrograph on the 2.7m Harlan J. Smith telescope of McDonald Observatory.  We used these data to study the distribution of the lines of the $\mathrm{d} ^3\Pi_g - \mathrm{a} ^3\Pi_u$ C$_2$ (Swan) bands. We show that the data are best represented with two rotational temperatures, with the lowest energy lines being at a relatively cool temperature and the higher energy lines being at a higher temperature.  We discuss the implications of this two temperature distribution and suggest future work.

\end{abstract}

\section{Introduction}

The processes by which the Solar System formed, and the conditions during its formation, are fundamental to understanding our existence and where extraterrestrial life could survive. Small planetesimals formed throughout the solar nebula, many of which were incorporated into the planets. For those that did not go into planet building, most were eventually ejected from the nascent Solar System, but some planetesimals remained. The predominantly rocky bodies are the present-day asteroids, while the icy bodies became the comets. Comets likely contain the least altered material from the primordial Solar System. Their low mass and generally large orbits result in little change during their lifetime. Therefore, cometary material is a unique probe into the early Solar System. Cometary compositional data can be compared to Solar System formation models and protoplanetary disk observations.

Comets are roughly a 50:50 mixture of ice and dust by mass. The ice is about 80\% H$_2$O, with smaller contributions from CO, CO$_2$, C$_2$H$_2$, H$_2$CO, CH$_4$, NH$_3$, etc.  The large amount of ice is an indicator that comets must have formed in the outer region of the solar nebula, beyond the water snow line.

As comets approach the Sun, the ice is heated and sublimes and the resultant gas flows outwards from the nucleus, forming the coma that we observe remotely.  The outflowing gas in the inner  coma is 
sufficiently dense that particle collisions affect the thermodynamics and chemistry of the gas \citep{RodgersCometsII}.  The outer edge of this collisional zone is the distance from the nucleus (cometocentric distance), r, at which the particle mean free path equals r. The value of r is dependent on the production rate of the gas and the lifetime of the molecule under study. For a comet such as Halley, with H$_2$O production rates above 10$^{29}$\,mol\,sec$^{-1}$, the collisional zone is several thousand kms. For the comets studied in this paper, the collisional zone is likely much smaller.
Outside of this collisional zone, material freely flows outward and only photochemical reactions happen. 

In this paper, we concentrate on high spectral-resolution observations in the optical.  Optical detectors tend to be more sensitive than other detectors, allowing observations of fainter comets.
Optical observations of comets provide direct and indirect information about cometary properties (e.g. composition, dust content, isotope ratios). In the optical, the molecular emission is mostly from fragments (daughters) of sublimated ices that have been photodissociated at least once. Hence, optical molecular emission is a secondary tracer of composition and a direct probe for the photochemical condition around the comet \citep{Jackson1973,Jackson1996,Gredel1989,Lambert1983}.

\section{Diatomic Carbon}

\begin{figure}
	\centering
	\includegraphics[scale=0.6]{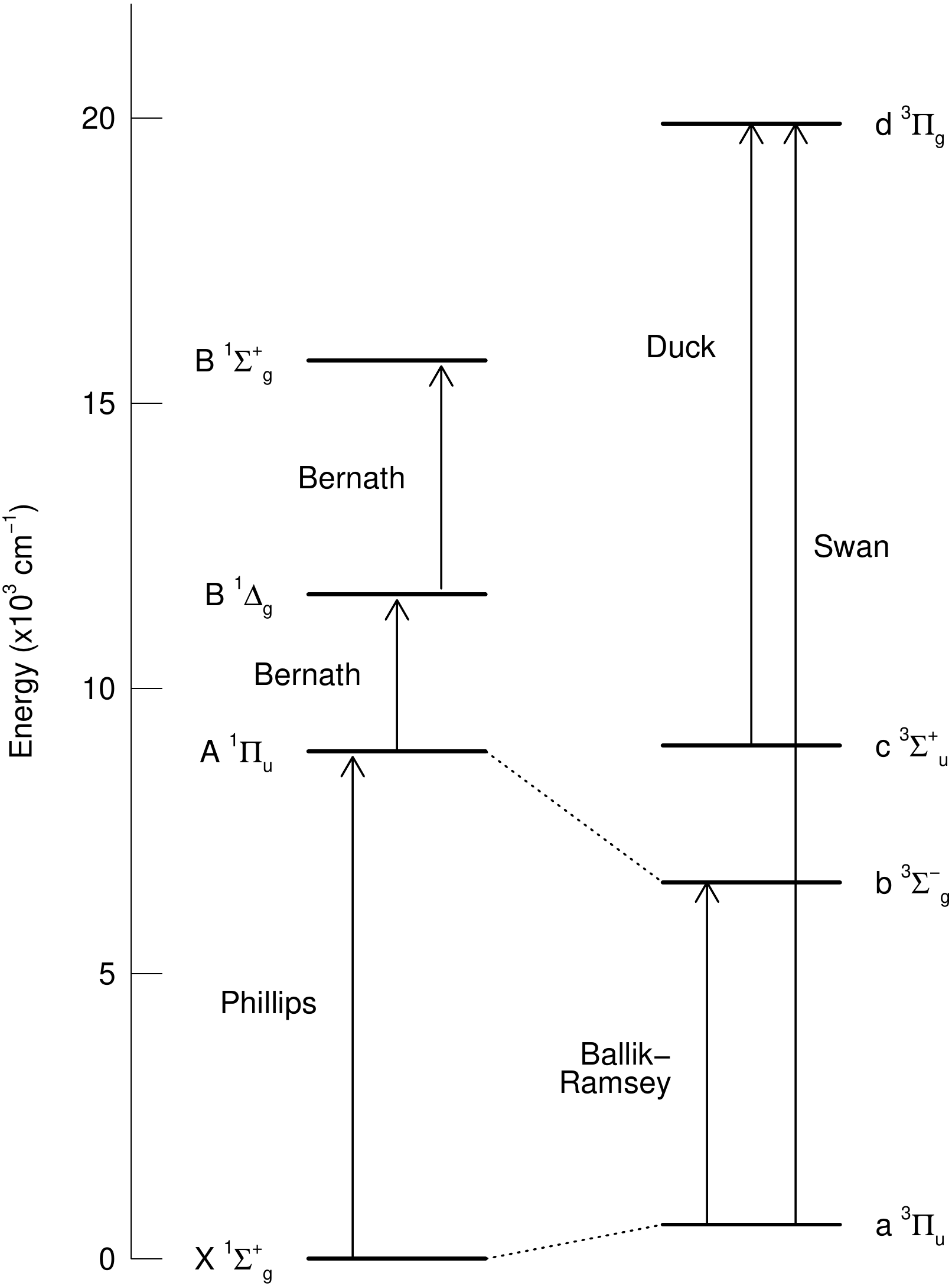}
	\caption{Shown are the first four singlet and triplet states for C$_2$.}
	\label{c2_transitions}
\end{figure}

The $\mathrm{d} ^3\Pi_g - \mathrm{a} ^3\Pi_u$ (Swan) bands of C$_{2}$ are very prominent features in normal (i.e. not carbon depleted) comets. As a result, C$_{2}$ emission has been used to measure production rates, isotope ratios, and as a taxonomic tool. C$_2$ is thought to be a daughter or, more likely granddaughter of some volatile (e.g. C$_2$H$_2$) \citep{Jackson1996, HonThesis}. Figure~\ref{c2_transitions}  shows the low energy electronic transitions. C$_2$'s lifetime against dissociation is $\sim 10^5$ seconds \citep{AnitaC2scale}. In almost all ground based observations, C$_2$ is in an optically thin gas which is non-collisional. Lacking collisions, C$_2$ reaches a fluorescence equilibrium with the Sun. 

As a homonuclear diatomic molecule, pure electric dipole rotational transitions are forbidden and only $\Delta\text{J}=0,\pm1$, are allowed for electronic transitions \citep{Herzberg}. Since $\Delta\Lambda = 0$ the Q branch quickly weakens; thus, for a given $J$, after one excitation cycle $J \rightarrow J,J\pm2$, with the rates of each being slightly different. Consequently, C$_2$ is ``pumped'' by the solar flux \citep{Herzberg, arpigny1965}.

Molecular models allow us to fit a rotational temperature to the spectra in order to characterize the physical processes. These are \underline{not} physical temperatures because most of the coma is non-collisional; they are convenient parameters that describe the emission as if the species were at the specified temperature. These temperatures arise from the competition of absorption of solar flux and emission from the molecule. With a series of laboratory or theoretical values for emission strength of transitions, vibrational/rotational temperatures are straightforward to quantify. This simplicity makes them useful for testing model predictions.

Adopting a simple resonance fluorescence model for C$_2$, it is expected the Swan bands would have a rotational temperature $\sim T_{\astrosun}$ \citep{C2_sun_temp}\footnote{\cite{Arpigny1967} shows that this is not quite accurate but it does not matter for our purposes.}. This would occur on a time scale of $\tau_{\mathrm{eq}} \sim \mathrm{n_{states}}R_\mathrm{abs}^{-1}$, where $\mathrm{n_{states}}$ is the number of rotational states and $R_{\mathrm{abs}}$ is the rate of absorption \citep{arpigny1964, Lambert1983}. The rate constant is the specific solar flux times the absorption rate of the lowest state. The lowest state is used because it has the lowest transition probability and would take the longest to equilibrate. For the (0,0) band, this timescale is about  $\sim 100$ seconds \citep{Lambert1983}.\footnote{The previous papers used A (the Einstein value) for the (1,0) band. The timescale varies inversely as A. Since A(0,0) is about 3x A(1,0), the (0,0) timescale is 1/3 the (1,0) timescale.} Observations by various investigators have found the rotational temperature varies inversely with heliocentric distance. \cite{Gredel1989} showed this can arise from intercombinational transitions between the triplet and singlet systems. With these additions, it is still expected that the above timescale of $\sim 100$ seconds is sufficient for the (0,0) band to reach steady state. 

\cite{Lambert1990} found a bimodal rotational temperature for the (0,0) Swan Band in spectra of comet 1P/Halley. The two temperature distribution persisted at different heliocentric and cometocentric distances; however, the values of the ``cold'' and ``hot'' components varied. Prior observations of C$_2$ in comets only found a single rotational temperature (see \cite{Nelson2018} for partial review). Similarly, only one temperature was seen by \cite{Lambert1990} when they observed an acetylene torch with the same set up. There has been little effort to examine whether this is a common feature of comets. A bimodal temperature distribution has been assumed for recent studies on C$_2$ \citep{Rousselot2012, Rousselot2015}.

Halley was a relatively bright comet, with high production rates of gas. With only one  example, we do not know if the bimodal temperature is unique to Halley or common in other comets. Our goal is to determine if this bimodal distribution is normal for comets, was due to Halley being such a productive comet, or if it was a function of factors such as heliocentric distance at the time of the observations. This paper details efforts to quantify the rotational temperatures of C$_2$ in additional comets. As will be discussed in Section~6, there are several current explanations for how the cometary C$_2$ would display two temperatures. These mechanisms can result in observable differences that can be probed with high resolution optical observations similar to the ones included in this paper.

\section{Observations and Reductions}
We observed comets 122P/de~Vico and 153P/Ikeya-Zhang using the Tull 2DCoude spectrograph on the 2.7\,m Harlan J. Smith telescope of McDonald Observatory \citep{TullCoude}. The spectra had a resolving power, R\,=\,$\lambda/\Delta\lambda$\,=\,60,000, using a slit that was 1.2\,arcsec wide and 8.2\,arcsec long. Table~\ref{obstab} is a log of the observations.  de~Vico was selected because it has almost no dust, making the initial analysis simpler. Ikeya-Zhang was selected because it was at a similar heliocentric distance to de~Vico but required us to develop methods for handling the more typical solar reflection spectrum.

\begin{table} [h]
	\centering
	\caption{Log of Observations}
	\begin{tabular}{c | c | c | c}
		Comet &  Date (UT) & r$_h$ (AU) & $\Delta$ (AU) \\
		\hline
		122P/de Vico & 1995 Oct 03 & 0.66 & 1.00 \\
		122P/de Vico & 1995 Oct 04 & 0.66 & 0.99 \\
		153P/Ikeya-Zhang & 2002 Apr 22 & 0.92& 0.42\\
		
	\end{tabular}

\label{obstab}
\end{table}

Preliminary data reduction followed a standard path.  The bias was removed from each spectrum using both the overscan region and bias frames.  A flat field was then removed from each spectrum and the echelle orders were defined in shape and width.  These definitions were used to extract the 1D spectra of the comets.  Wavelengths for each pixel in the extracted spectra were defined using observations of a ThAr lamp and had an rms error of the wavelength of $\sim24$\,m\AA.
The spectra were Doppler shifted to the laboratory rest frame using the geocentric radial velocity of the comet.  In addition to the comet observations, we obtained observations of the daytime sky using a ground-glass solar port that allows the sky to be imaged through the slit of the spectrograph and to follow the normal light path of the instrument.

The Swan band sequences are spread over many echelle orders, each of which has a slightly different sensitivity. This was corrected by dividing a separate solar spectrum obtained through the solar port with the Kurucz Solar Atlas data \citep{Kurucz} and then fitting a low order polynomial to this ratio over each order. Dividing the comet observations by these curves removes most of the inter- and intra-order variation. Next, all the relevant orders are combined into one spectrum. This puts the whole band on the same relative flux scale, but is not an absolute calibration.

\begin{figure}
	\centering
	\includegraphics[scale=1]{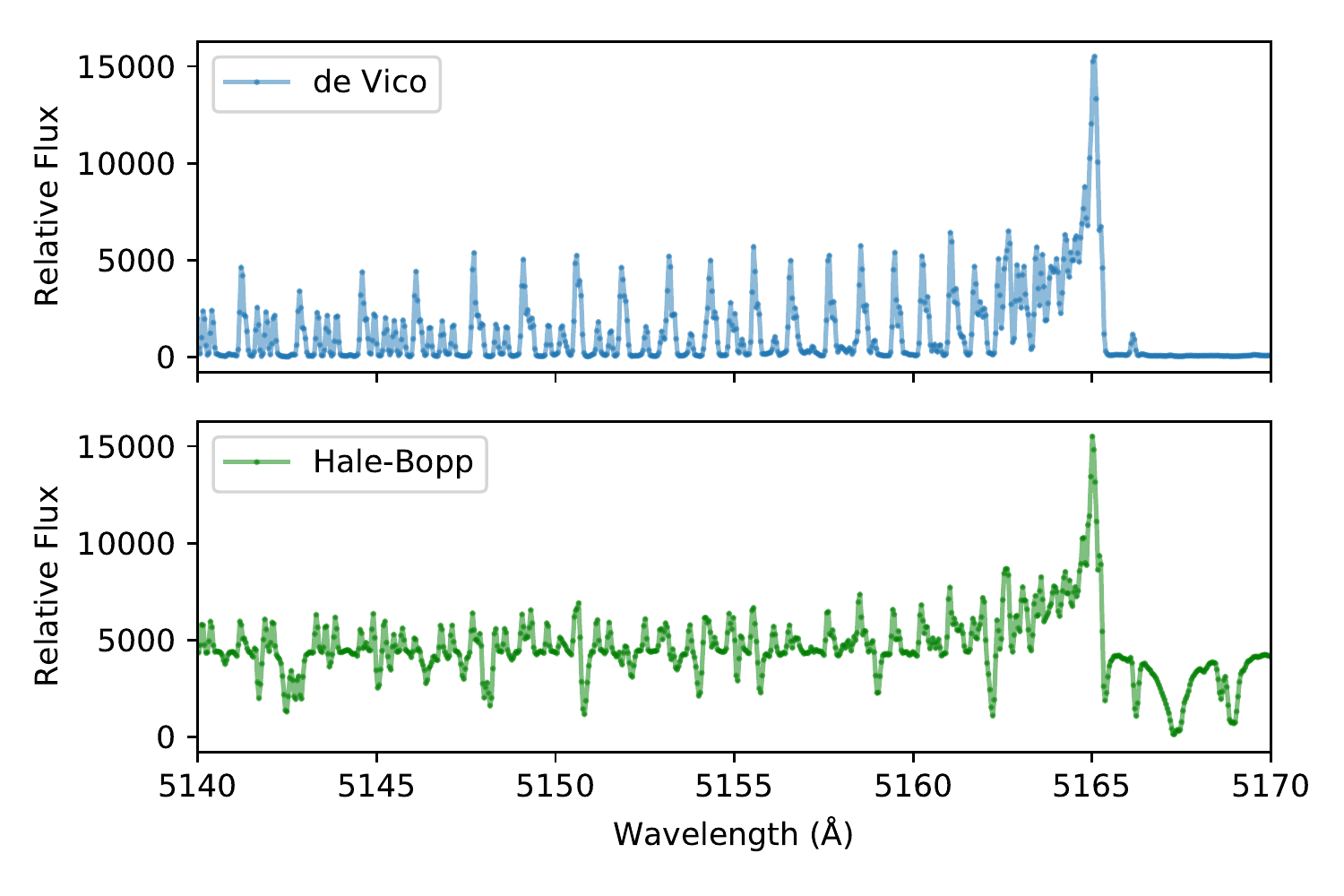}
	\caption{The top and bottom spectra highlight the large variation in the reflected dust component between comets. Absorption features are mostly absent in de Vico, while they are $\sim$50\% of the C$_2$ (0,0) bandhead for Hale-Bopp.}
	\label{dusty}
\end{figure}

In the optical, dust serves to reflect the solar spectrum. This reflection may have a color different than the Sun, causing the spectrum to appear tilted. The relative strength of the reflected solar spectrum compared to the molecular emission varies by comet, as shown in Figure \ref{dusty}. With the low dust contamination of de~Vico, dust removal only slightly changes the models.
For Ikeya-Zhang, removing the dust is necessary to model properly the underlying C$_2$ spectrum. Removing the dust amounts to estimating a continuum and identifying solar absorption features.

\begin{figure}
	\centering
	\includegraphics[scale=1]{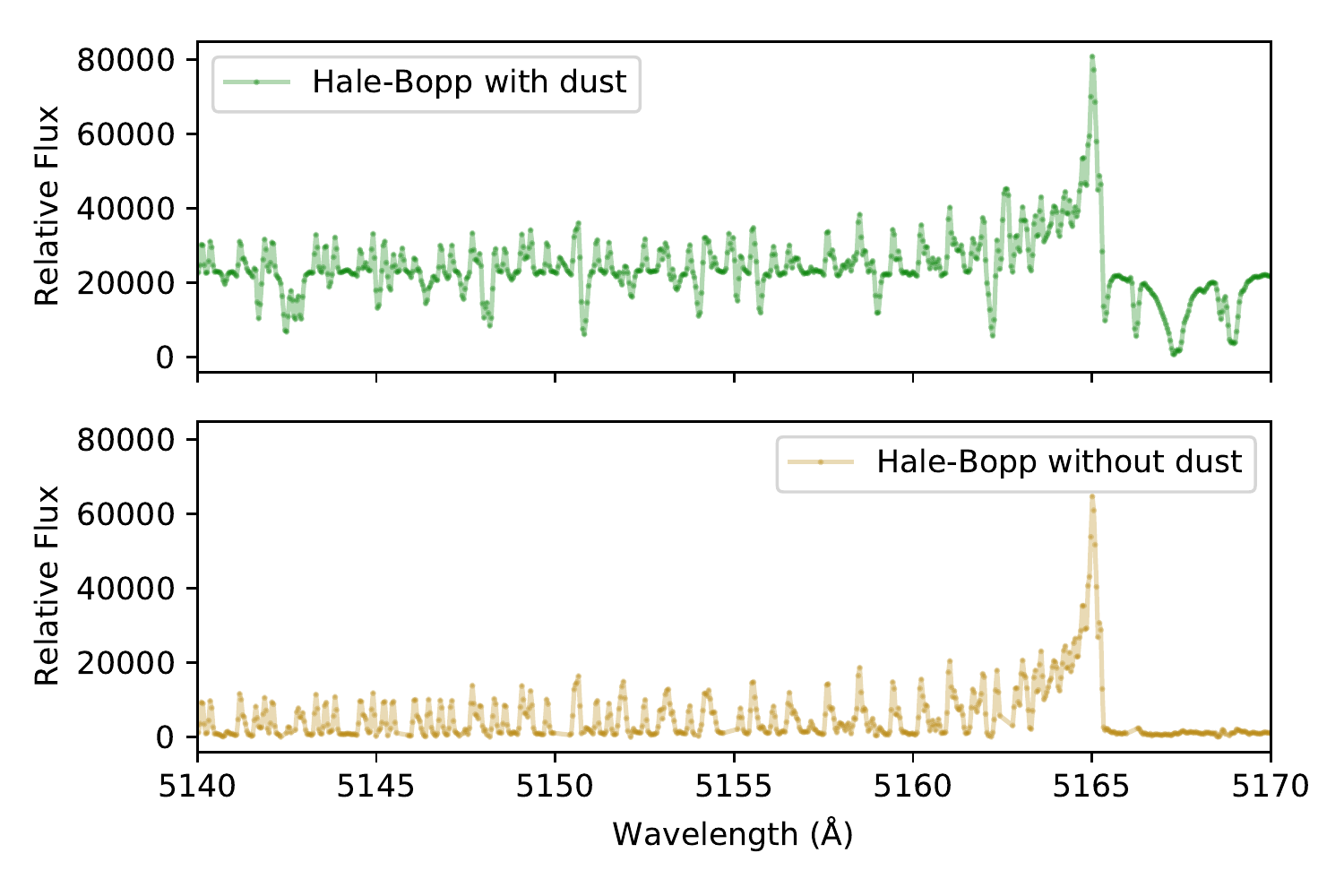}
	\caption{Before and after subtracting the dust component from Hale-Bopp's optocenter.}
	\label{dustStuff}
\end{figure}

To subtract the dust, we first removed the tilt introduced by the dust color. Then we scaled the solar spectrum from the solar port (or, if unavailable, the Kurucz Solar Atlas) to the data based on absorption features that were also seen in the cometary data. The results for a very dusty comet, C/1995~O1 (Hale-Bopp) are shown in Figure \ref{dustStuff}. This region is representative of the procedure applied over the entire band. As can be seen from this figure, we can remove the dusty continuum very accurately.


Contaminating emission from other species, such as NH$_2$, was masked out by using the atlas from \cite{devicoAtlas} and comparing reasonable model spectra to the data. A line was masked out if a large deviation was observed between the model and data (e.g. one spin component being anomalously strong) and a contamination was likely to blend into a C$_2$ emission feature.  This masking was done after the dust removal described above. 

%


\section{Rotational Temperature}
 \cite{Lambert1990} found a bimodal rotational temperature for the (0,0) Swan Band in spectra of comet 1P/Halley.  The  J-level lines below $\sim15$ showed a much lower rotational temperature ($665\pm70$ K) than the higher J-level lines ($3330\pm170$ K).  We performed a similar analysis with de~Vico and Ikeya-Zhang to see if other comets have a bimodal temperature. In addition, our data cover a much larger wavelength range (4880--5166 \AA\ for our data; 5131--5166 \AA\ for Lambert et al.), and therefore a greater range of J values (45 vs 84). 

In general, determining the energy distribution for a system of molecules is a difficult task. In the special case of thermal equilibrium, this energy distribution is only a function of the temperature 
\begin{equation} 
X_m = \frac{X}{U}g_m\exp\left(-\frac{E_m}{kT}\right)
\end{equation} where the subscript $m$ denotes quantities belonging to that state, $X$ is the total number of molecules, $U$ is the partition function, $g$ is the degeneracy of the state, and $E$ is the energy, and $T$ is the temperature. This parameterization is a good starting point for examining other systems as well because it can be used as a metric for the population function. 

For an optically thin, non-collisional gas dominated by spontaneous emission, the flux of a transition goes as 
\begin{equation}
F_{mn} \propto X_m E_{mn} A_{mn}
\end{equation}
where $m,n$ are indices for the upper and lower state and A$_{mn}$ is the Einstein value for that transition. Let $F_\lambda$ be the integrated flux of an observed line centered at a wavelength $\lambda$. Taking the population function as a Boltzmann distribution gives

\begin{equation}
F_\lambda = \sum_i F_i \propto \sum_i \frac{S_i}{\lambda_i^4}\exp\left(-\frac{E_i^\prime}{kT}\right)
\label{eq:Boltz}
\end{equation}
 where $i$ indexes over the transitions which are unresolved at this wavelength, $S_i$ is the line strength of a particular transition, $E_i^\prime$ is the energy of the excited state, $\lambda_i$ is the wavelength of the transition, and $T$ is the temperature. $S_i$, $\lambda_i$, and $E_i^\prime$ are all known from laboratory work, so temperature is the only variable remaining. One can then create a ``Boltzmann plot" to determine the trend of $\log(F_\lambda/[S_i/\lambda^4])$ with energy. The lines fit to these data points then have a slope of $-1/kT$. 
 
 Equation \ref{eq:Boltz} is used because many of our lines are unresolved for a variety of reasons.
 We fit for temperature to see if there was one or more temperatures; we used all unblended lines but used blended lines only if the excitation energy was the same for the blended lines. For blended lines, we replaced $E_i^\prime$ with an average over the blended states. This usually occurred with P (or R) branch lines which have the same $N^\prime$. There are several blend cases which do not meet the energy requirement so we did not use them in this analysis. These cases include: 1) blends between low and high P branch components around the bandhead, 2) blends between P and R branches, 3) blends between different bands (e.g. (0,0) and (1,1)). 

We use line strengths and energies from a PGOPHER model of the Swan bands (\cite{Colin2017, Brooke2013};  see below for description).
Each emission line,
composed of one or more C$_2$ lines, is considered a separate flux value to be fit. The user identifies the approximate locations of all large transitions for an observed blended line.  This is straightforward using a Fortrat diagram for C$_2$. 

We use weighted least squares to find the best fit. Let $b_i$ denote the number of transitions for blend $i$. For blend $i$, we attempted to fit the flux as

\begin{equation}
	f(\lambda, \boldsymbol{A},\boldsymbol{\mu}, s) = \frac{1}{\sqrt{2\pi\sigma^2}}\sum_{j}^{b_i} A_j \exp\left(-\frac{\left(\lambda - \mu_j - s\right)^2}{2\sigma^2}\right)
	\label{flux_fit}
\end{equation}

\noindent where $\lambda$ is wavelength, $\boldsymbol{\mu}$ are the transition locations, $A_j$ is the integrated flux for the jth component, and $s$ is an empirical wavelength shift to handle wavelength calibration errors. If the normal matrix is linearly independent, the form above is correct. If not, we have to revise our fit to reflect the inability to separate transitions. 

To determine whether the system is linearly independent, we calculated the eigenvalues for the normal matrix. The lowest eigenvalue is discarded because it comes from the empirical shift term. We count the number of eigenvalues below a threshold\footnote{This threshold was determined by testing on many lines and appears to be adequate.} of 10$^{-3}$, $d$. These are considered to be linearly dependent. For this blend, the code finds the first $d$ closest transitions and replaces them with merged transitions.

We keep track of the lines which contribute to a merged transition. Their line center is assumed to be the average of these underlying lines because the lines have already been required to have similar strengths. This merging process only occurs when the wavelengths are very similar, generally when the transitions are separated by less than $\Delta \lambda$ between adjacent pixels. By construction, these components will not exhibit a noticeable difference in broadening. The line fitting process is redone with these revisions. Without the merged transitions, the $A_j$ correspond directly to integrated flux values. With the merged transitions, $A_j$ are the sum of all transitions which belong to that unresolved blend.

Finally, all successful line fits\footnote{A fit was successful if the covariance matrix was positive semidefinite. We evaluated this using a Cholesky decomposition.} are transformed as above and fit using the piecewise linear function, pwlf \citep{pwlf}, python package. We made minor modifications to this code so we could use covariance matrices from the line fits. For the Boltzmann plots, we used the following function
\begin{equation}
g(E) = g_0 + E\left(a_1 + a_2u(E-b)\right)
\label{beqfit}
\end{equation}
\noindent where $g_0$ is the y intercept, $a_1$ and $a_2$ are related to temperatures as $T_i = \left(\sum_{j\le i} a_j\right)^{-1}$, and $u(E-b)$ is the unit step function offset by the break point $b$ (the break point is the energy level at which our fits transition from one temperature to another). For propagation of error, we approximate the piecewise behavior at a value $b$ with the logistic function: $u(E-b) \approx \left(1 + \exp[-2t(E-b)]\right)^{-1}$, where larger values of user-defined constant $t$ cause a sharper transition\footnote{After trial-and-error we found $t=10^{-2}$ was sufficient for our analysis}. The fitting routine tries to find the optimal break point for the transition from the colder to the warmer temperature.  We place no restrictions on where that can be and, in cases where a single temperature fit is as reasonable as two temperatures, this break point can end up at seemingly random energy levels.  This probably confirms that a second temperature is not needed. 



\subsection{Error Analysis}
\added{As discussed below, the temperatures are fairly sensitive to the treatment of the data and estimation of uncertainties. Therefore we provide an outline of the sources of error considered in this analysis. Our analysis has considered the following effects (in this order): 1) the blaze function, 2) residual large scale sensitivity changes between orders on the spectrograph, 3) the dust color, 4) sensitivity changes over a single order in the spectrograph, 5) the effect of combining orders with overlapping wavelength sections together, 6) sky color in the solar port, and 7) scaling the solar observation to match the reflected solar spectrum in the comet.} 

\added{Adequate estimation of the continuum was important for many of these steps. We assumed that the most common flux value over some contiguous wavelength range (i.e. the mode) was the continuum. Our implementation uses a sliding window mode. The windowing controls the amount of smoothing over a wavelength region, so it must be large enough to easily encompass most spectral features. The sliding amount controls the rate of change between adjacent mode estimates, with smaller values creating a smoother distribution of continuum estimates. To estimate the mode on each of these wavelength subsets, we apply a smoothing kernel to remove small scale scatter in the flux around the `real' continuum value.}

\added{There are several effects which are not accounted for in our error analysis. These are: 1) influence of prior reduction steps, 2) continuum estimation parameters (i.e. window width, sliding length, smoothing kernel size), 3) the effect of interpolation rather than resampling onto different $\Delta \lambda$ spacing between orders, 4) wavelength errors, 5) broad absorption lines. The wavelength errors were handled with the empirical shift term in Equation \ref{flux_fit} but not in our error propagation. We experimented with different kernel sizes, settling on one which provided enough smoothing that the resulting flux distribution only had one peak. To compensate for this smoothing kernel, we take the FWHM/2.35 of the flux distribution as the uncertainty of a mode estimate. We have not fully explored the influence of either the window width or the sliding length. Observations of Hale-Bopp were used to find good starting point values for these parameters because Hale-Bopp is very dusty. For comets at low SNR, using a smaller sliding length and wider window were needed to extract the dust continuum. Finally, to correct for order sensitivity changes, we assume that the windowed mode properly traces the continuum. This assumption breaks down near H$\beta$ because it spans most of an echelle order. Our heuristic identifies H$\beta$ as a continuum feature and attempts to correct it, imprinting an erroneous emission feature. We have mitigated the influence of this issue by order correcting from red to blue (to prevent the overlap between orders from being unduly affected), excluded lines close to the core (which are likely affected by the Swings effect \citep{Swings1941} as well), and because our band is blue degraded, relatively low amounts of emission occur here.}

\section{Results}
\subsection{Boltzmann Plots}
\begin{figure}
	\centering
	\includegraphics[scale=0.6]{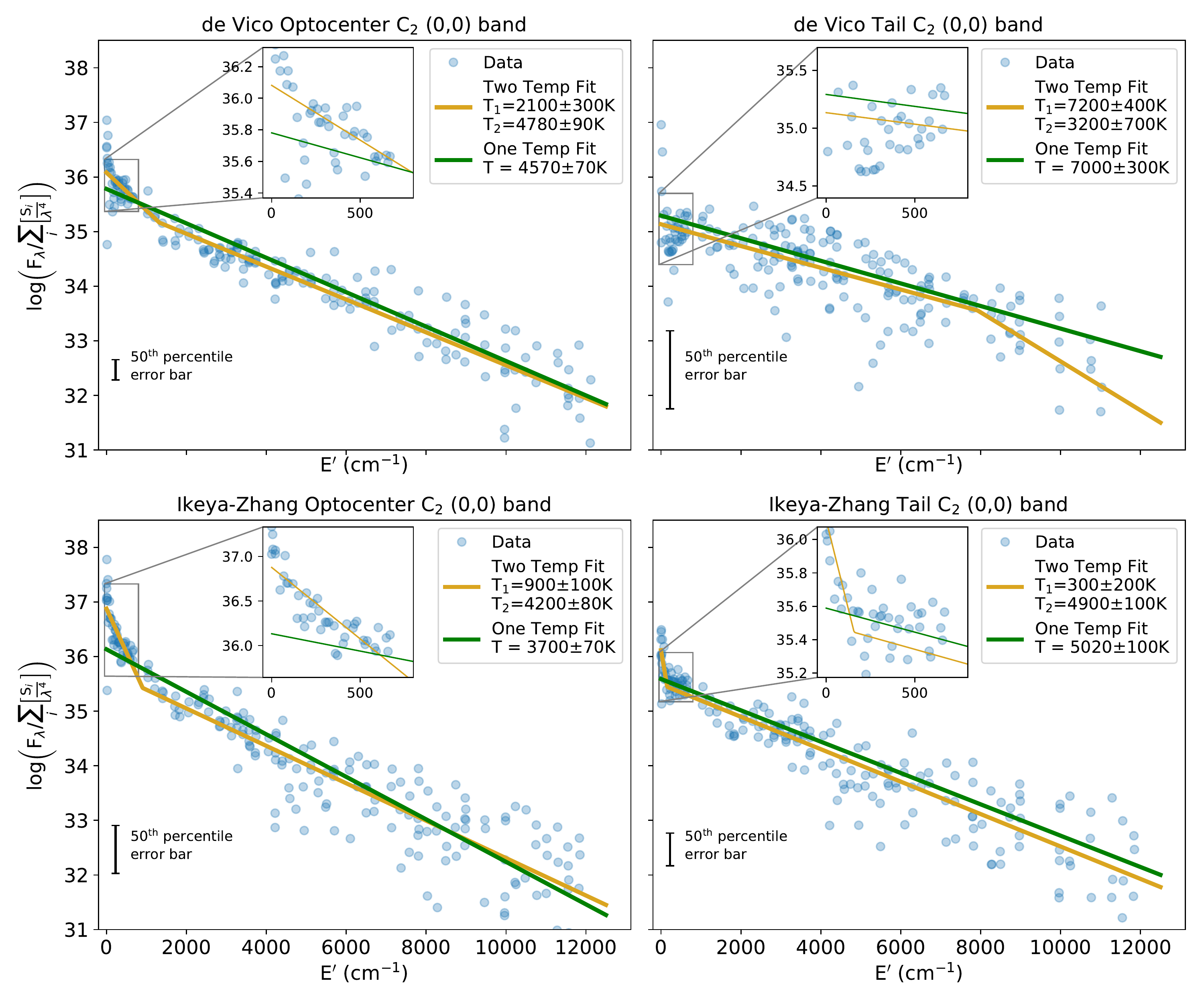}
	\caption{Boltzmann plots for de Vico (top) and Ikeya-Zhang (bottom). The left column are observations of the coma and the right are observations in the tail. These values have been shifted so they all occupy the same y-value range. The optocenter observations are well fit by a two temperature trend. Whether the tails also have two temperatures is not obvious from the plots.}
	\label{bplot00}
\end{figure}

\begin{figure}
	\centering
	\includegraphics[scale=0.6]{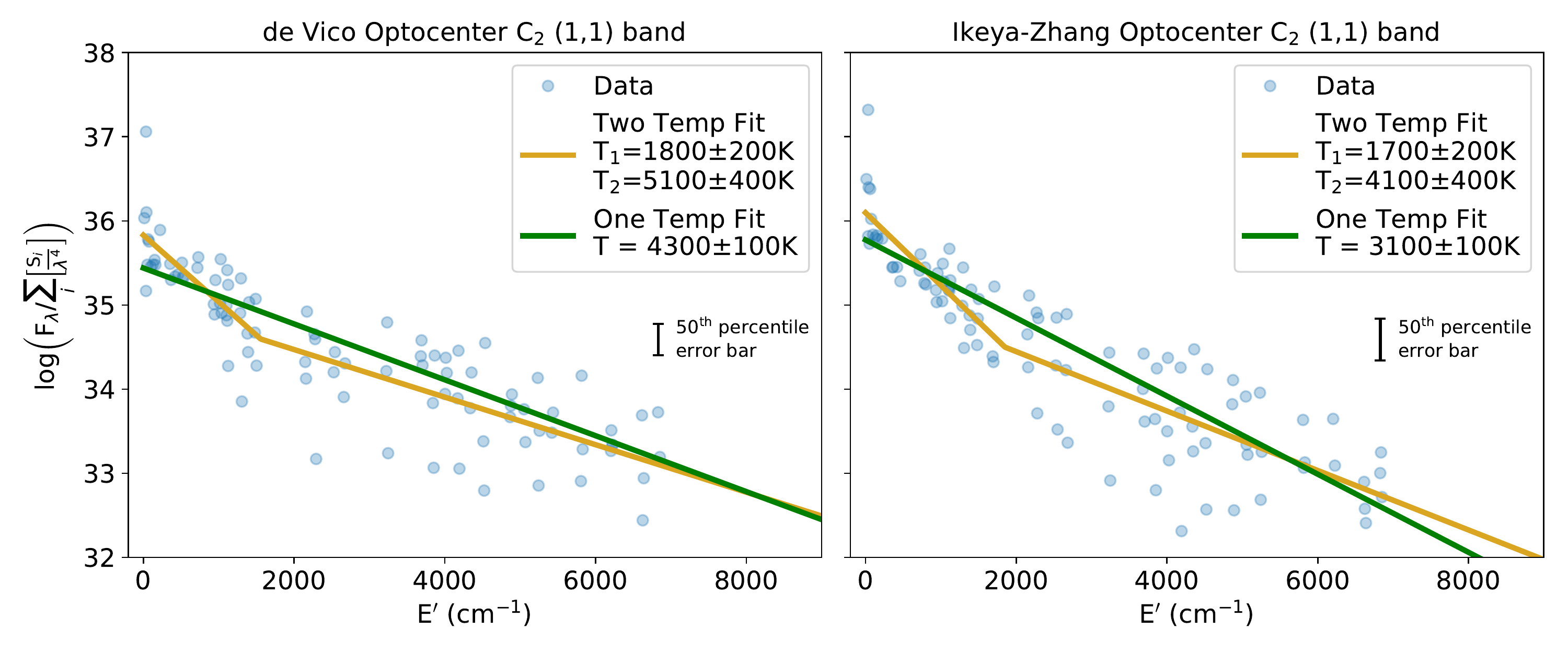}
	\caption{Boltzmann plots for the (1,1) band in de Vico (left) and Ikeya-Zhang (right). Qualitatively these plots agree with the (0,0) Boltzmann plots.}
	\label{bplot11}
\end{figure}

The Boltzmann plots for the (0,0), and (1,1) bands for de~Vico and Ikeya-Zhang are shown in Figures \ref{bplot00} and \ref{bplot11} respectively. Our observations included spectra of the optocenters of the comets as well as spectra in the tail 72,500 km from the optocenter for de~Vico and 28,000 km from the optocenter for Ikeya-Zhang. The Boltzmann plot energies shown are relative to the ground energy, around 19860\,cm$^{-1}$. A summary of all Boltzmann fits is given in Table \ref{btable}.  The magnitude of the correlation coefficients for the two temperatures were all less than 0.08, so we do not report them here. \added{The scatter in line fluxes is accounted for in the temperature uncertainties by rescaling the covariance matrix derived from Equation \ref{beqfit} by the $\chi^2_\nu$.} The plots for the (0,0) optocenter strongly suggest there are two temperatures and this is corroborated by the large improvement in $\chi^2_\nu$ and Bayesian Information Criterion (BIC) in the two temperature fits relative to the one temperature fits. We are not convinced that two temperatures are needed to fit the tail data for either comet. The ``heating'' of C$_2$ can be seen by comparing the optocenter and tail observations for the (0,0) band. The (1,1) fits in the tail follow the same trend as the (0,0); however, they have so few points, from the low SNR, that we do not include them in this discussion.

Based on the arguments outlined in the introduction, where we expected a maximum rotational temperature of $\mathrm{T}_{\astrosun} \sim 5780 \ \mathrm{K}$, we do not believe the temperatures for de Vico's tail are real. We have included it to highlight shortcomings of this method and possible issues with our analysis. Since the temperatures go as 1/slope of the best fit lines, and these slopes are of order $10^{-4}$, small deviations in the best fit slope can dramatically alter the resulting temperature value. For the de Vico tail data, the difference in the best fits are almost imperceptible on the plot, but result in a 200 K shift between $T$ and the ``hot'' temperature! Consequently, the accuracy of the integrated fluxes has a large impact on the fit temperatures. However, we believe our determination of whether one or two temperatures are present is robust and, when two temperatures are present, the large difference in temperature between the low and high temperatures is real. 

\begin{figure}
	\centering
	\includegraphics[scale=0.8]{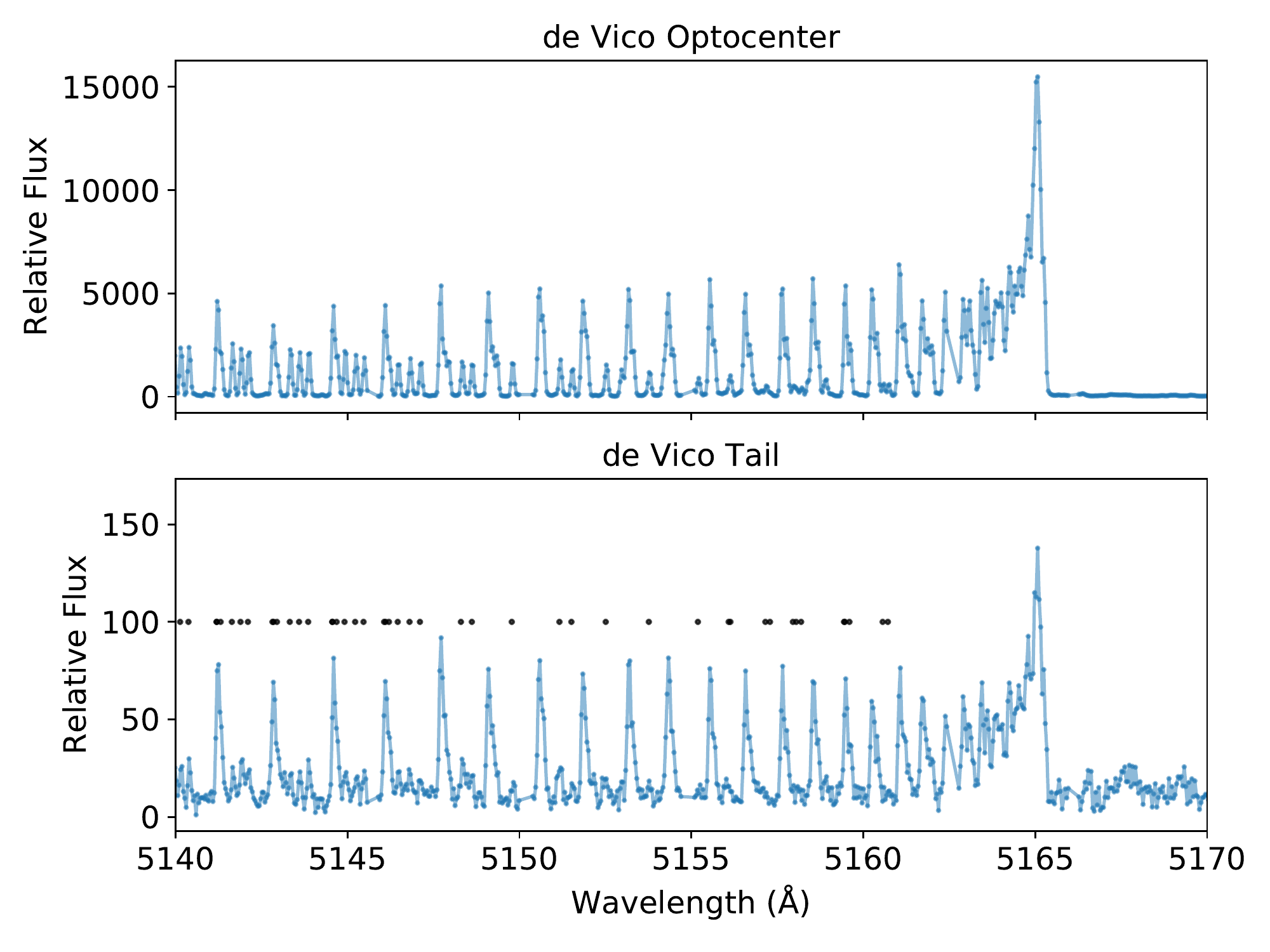}
	\caption{Comparison of de Vico's optocenter and tail spectra. The black dots in the lower panel indicate the transitions we attempt to fit in this region.}
	\label{tail_noise}
\end{figure}

de Vico's tail has the lowest SNR of all the data presented. Figure \ref{tail_noise} displays a portion of the optocenter and tail data for de Vico after the reductions. The black dots in the lower panel of Figure \ref{tail_noise} are positions of lines we tried to fit. The bandhead signal differs by a factor of 100 between the two positions. By eye, there is a residual continuum for the tail of $\sim 20-25$ that results from the more noisy data. Hence, for many lines, the continuum becomes a major source of error in low signal data. This is particularly troublesome for the low J transitions since these have a weak signal already. Even in the simplest case, where the continuum has a constant offset, the effects on the Boltzmann plot are nonlinear because it uses the log of the flux. All fit line fluxes have had a continuum correction term subtracted from them to crudely account for the extra flux. The residual continuum level appears roughly constant between de Vico and Ikeya-Zhang. While we expect the continuum to become important when it is of the same order as the signal, Ikeya-Zhang's tail has twice the signal of de Vico's and already the values for the best fit are much more reasonable. 


We tested the influence of break points ($b$ in Equation \ref{beqfit}) on our fits. We compared the variable break point fits to models where the location of the break point was specified. The best fit temperatures from using different breakpoints agreed within the errors. There was a clear fit minimum for break points at the value fit automatically by our routine. Bad flux fits or misidentifying lines are also important sources of error on the temperatures. The low temperature fits have fewer points, so they are more adversely effected by misidentification and bad fluxes.

With those caveats in mind, there are a few comparisons we can make between the Boltzmann plots and their best fits. de Vico has a higher ``hot" temperature, regardless of whether the single or two temperature model is adopted. The fits improved by at least 30 percent when two temperatures are suggested. The fits do not require two temperatures for the tail data, while all optocenter data show two temperatures.

We have limited our analysis to the (0,0) and (1,1) bands despite having data for many more bands. This was done for several reasons. First, there is not theoretical motivation to expect the other bands to have qualitatively different temperature(s). The Einstein A values for the bright Swan emission bands are all of the same order of magnitude. The other reason concerns the accuracy of the theoretical models for the Swan system. These models are largely based on the \cite{Tanabashi2007} lab work. The models are very good at replicating our observations in the (0,0) band. There is noticeable divergence in wavelength at high J ($\ge 80$) levels, likely because these lines are difficult to create in the lab. However, the intensities remain fairly accurate well beyond these J levels. This situation is worse for other bands, with the divergence of (1,1) happening for J $>$50. The experimental data for the other band sequences often did not reach beyond the J = 40 level. These problems can partially be mitigated using other empirical line lists for the labeling stage. 
\begin{table}
	\centering
	\caption{Summary of Boltzmann fits for de Vico and Ikeya-Zhang.}
	\begin{tabular}{c | C | C | C | C}
		& \mathrm{T}_1 \ (\mathrm{K}) & \mathrm{T}_2 \ (\mathrm{K}) & \chi^2_\nu & \mathrm{BIC} \\
		\hline
		de Vico (0,0) & 4470 \pm 70 & - & 6.7 & 1290 \\
		Optocenter & 1500\pm 300 & 4890 \pm 80 & 3.7  & 716 \\
		\hline
		de Vico (0,0) & 6600 \pm 300 & - & 2.7 & 459 \\
		Tail & 7000 \pm 400 & 2800 \pm 600 & 2.8 & 490 \\
		\hline
		de Vico (1,1) & 4000 \pm 100 & - & 4.6 & 437 \\
		Optocenter & 1800 \pm 200 & 5400 \pm 500 & 3.4 & 323 \\
		\hline
		Ikeya-Zhang (0,0) & 3550 \pm 60 & - & 6.7 & 1338 \\
		Optocenter & 1260 \pm 100 & 4300 \pm 100 & 3.5 & 713 \\
		\hline
		Ikeya-Zhang (0,0) & 4950 \pm 100 & - & 2.2 & 433 \\
		Tail & 500 \pm 100 & 5100 \pm 100 & 2.3 & 461 \\
		\hline
		Ikeya-Zhang (1,1) & 2700 \pm 100 & - & 6.0 & 535 \\
		Optocenter & 300 \pm 190 & 3200 \pm 200 & 4.2 & 375 \\

	\end{tabular}

\label{btable}
\end{table}

\subsection{PGOPHER}

While the Boltzmann plots are an excellent tool for identifying that the (0,0) band C$_2$ lines are best fit by two different rotational temperatures, it makes the assumption that all of the C$_2$ can be described by a Boltzmann gas distribution.  We also tested the reasonableness of the two temperature fit by modeling the data using a molecular spectral simulation code, PGOPHER.

As described in \cite{Colin2017}, PGOPHER is a general purpose program for simulating and fitting spectra. PGOPHER calculates line strengths by solving the Hamiltonian for the molecule of interest. To simulate a spectrum, PGOPHER also requires a population function to reflect the experimental conditions. It fully or partially automates the processes of determining spectroscopic constants. To do this, it allows for great flexibility in setting up the system to model. This includes arbitrary perturbations, complex or custom partition functions, energy levels, etc. \cite{Brooke2013} used PGOPHER to fit spectroscopic constants to the experimental Swan band data from \cite{Tanabashi2007}. These values are the current standard for the C$_2$ Swan System. 

PGOPHER can also work to solve experimental conditions if the molecular constants are known, flipping the original paradigm around. If the model is sufficient, then we can simulate and fit our expected temperature function to the data. This was the approach of \cite{Nelson2018}. In this paper, we have added PGOPHER fits of the data for Ikeya-Zhang.  The results for the two comets are shown in Table~\ref{pgoTab}. We have addressed some possible sources of these large errors in \cite{Nelson2018}. We think there are three important sources of error for our data, 1) wavelength shifts either from the calibration or unaddressed perturbations in the model, 2) continuum levels, since these will drive residuals outside of emission lines, and 3) high SNR -- the better the data, the more obvious there are problems with the model. This last point is why de Vico, our best data set, seems to fare the worst. 
\begin{table} [h]
	\centering
	\caption{Best Fit parameters for the $\Delta v=0$ band sequence in de Vico and Ikeya-Zhang }
	\label{pgoTab}
	\begin{tabular}{c | C | C | C | C | C}
		  & \mathrm{T}_1 & \mathrm{T}_2 & a & \mathrm{T}_v & \chi^2_\nu\\
		\hline
	de Vico  & 3592(49)& - & - & 5223(66)& 2639\\
	Optocenter & 634(91)& 4240(100)& 0.139(18)& 5015(63) & 2543 \\
	\hline
	de Vico  & 5630(130)& - & - & 6110(110)& 37.8\\
	Tail & - & - & - & - & \mathrm{Did \ not \ Converge} \\
	\hline
	Ikeya-Zhang & 3233(39) & - & - & 4717(55) & 320 \\
	Optocenter &  337(28) & 3818(63) & 0.172(11) & 4402(56) & 292\\
	\hline
	Ikeya-Zhang & 3818(56) & - & - & 5857(85) & 312 \\
	Tail &  - & - & - & - &  \mathrm{Did \ not \ Converge} \\
	
	\end{tabular}
\tablecomments{`a' is the relative population between T$_1$ and T$_2$}

\end{table} 

\begin{figure}[h]
	\centering
	\includegraphics[scale=0.6]{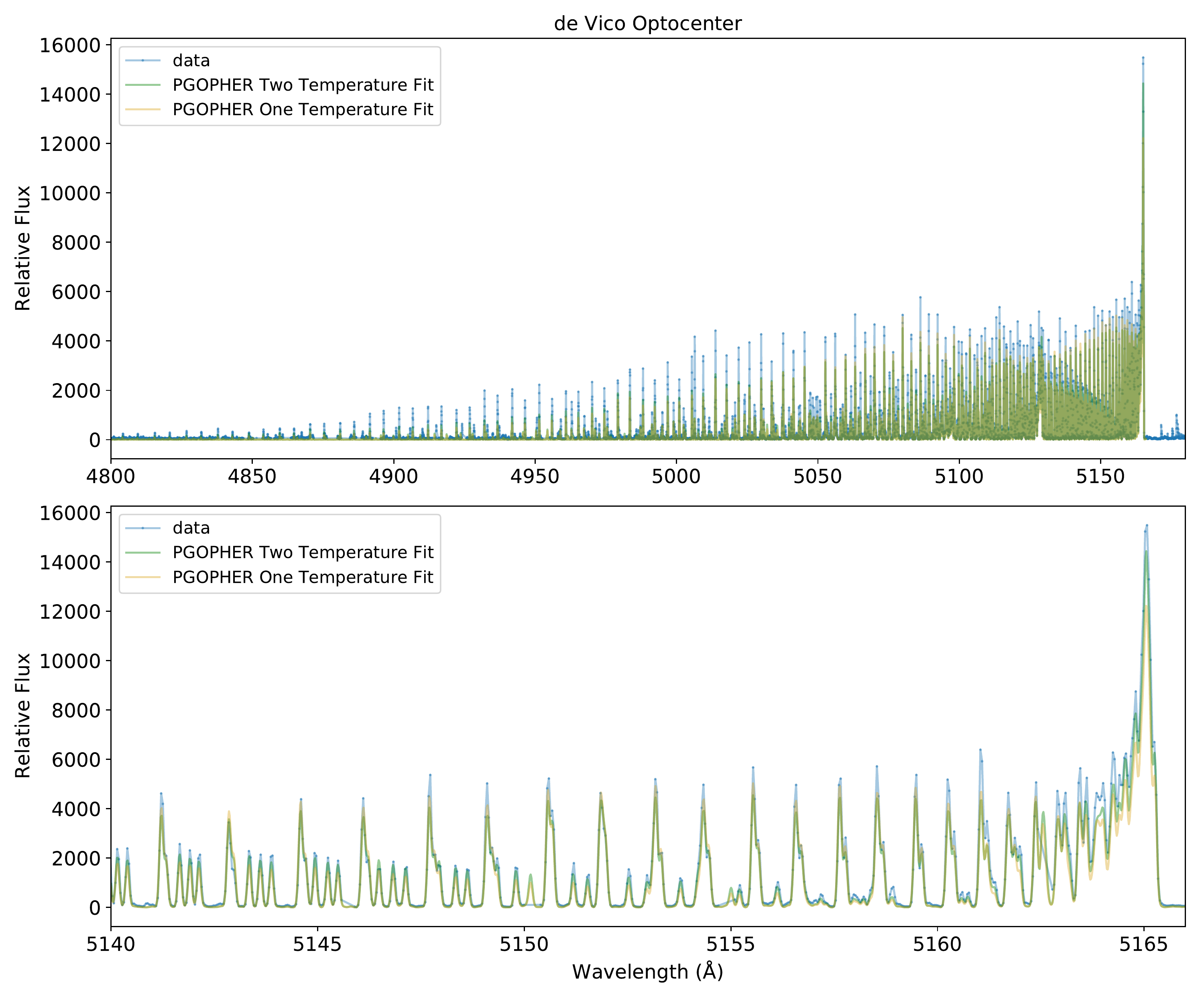}
	\caption{Both panels are comparisons of the de Vico optocenter observations with the best fit PGOPHER models from Table \ref{pgoTab}. The top panel displays most of the $\Delta v = 0$ band sequence and the bottom panel focuses on the emission around the (0,0) bandhead. The PGOPHER models have small deviations in the 5130 - 5166 \AA \  region. Many transitions blue of 5130 \AA \ are underestimated by the models.}
	\label{dev_pgo}
\end{figure}

\begin{figure}[h]
   	\centering
   	\includegraphics[scale=0.6]{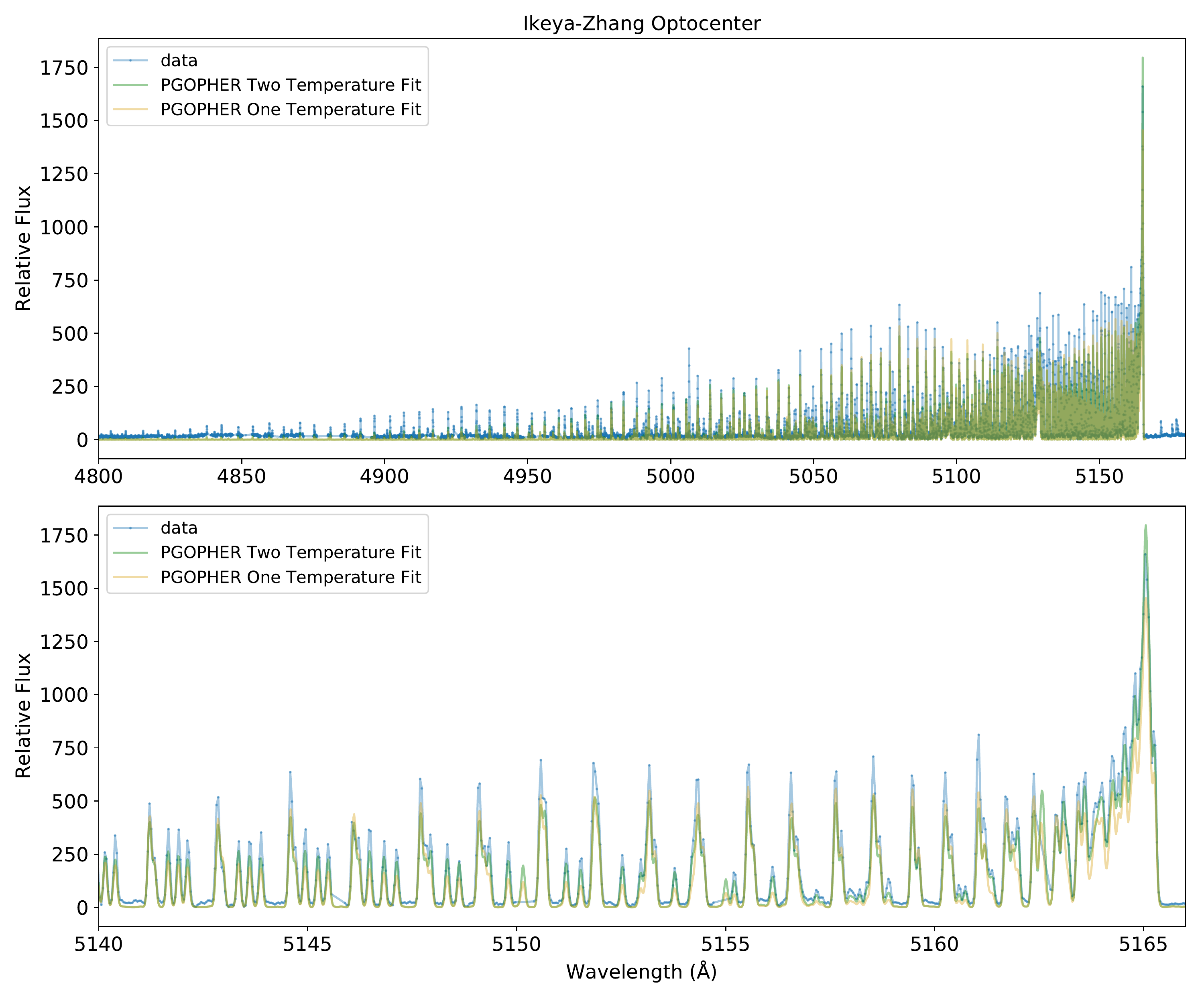}
   	
   	\caption{Both panels are comparisons of the Ikeya-Zhang optocenter observations with the best fit PGOPHER models from Table \ref{pgoTab}. The top panel displays most of the $\Delta v = 0$ band sequence and the bottom panel focuses on the emission around the (0,0) bandhead. Compared to de Vico in Figure \ref{dev_pgo}, the model appears worse in the 5140 - 5166 \AA \ region while the mismatches blue of 5130 \AA \ are not as pronounced.}
   	\label{iz_pgo}
\end{figure}

These PGOPHER fits agree mostly with our Boltzmann plots. The non-converging fits for the tails had very small $a$ values, where a is the ratio of the percentage of gas with temperature 1 to temperature 2. The fact that de Vico is hotter than Ikeya-Zhang agrees with the expected $r_h$ dependence. The de~Vico fits seem to be 500-600 K off from the PGOPHER fits, whereas there is perhaps a 500-1100 K offset for the Ikeya-Zhang fits. While we did not focus on the vibrational temperature, our results of $\mathrm{T}_v > \mathrm{T}_r$ agree with previous studies \citep{Lambert1983, Gredel1989}. \added{Figures \ref{dev_pgo} and \ref{iz_pgo} show the PGOPHER fits for the de Vico optocenter and Ikeya-Zhang optocenter respectively. The one and two temperature models perform similarly outside the bandhead for both comets. The zoomed-in Ikeya-Zhang data show a stronger preference for the two temperature model compared to de Vico, with noticeable discrepancies between all the low R branch lines shown. This qualitatively agrees with the relative improvements for these observations when using the Boltzmann plots. In both observations, the (0,0) bandhead is better matched by the two temperature model and, given its relative flux compared to the other lines, likely drives the fitting process. It is also apparent how much work remains to correctly capture the $\Delta v = 0$ emission for C$_2$ as many lines blue of 5130~\AA \ have underestimated flux. }

\section{Discussion}

We have used high spectral-resolution observations of two different comets to show that their C$_2$ emission spectra of the optocenter are probably represented best by bimodal rotational temperatures. This confirms what was found by \cite{Lambert1990} for comet 1P/Halley and suggests that this might be common for cometary spectra, at least for comets closer to the Sun than 1\,AU.

There are three contemporary explanations for how this bimodal temperature could arise: 1) The bimodal temperature is a result of C$_2$ in resonance fluorescence in a low density environment via intercombination and satellite transitions; 2) The two temperatures are from competing formation pathways from a parent molecule being photodissociated; 3) There are multiple populations of C$_2$ that are superimposed (e.g. an old population at steady state and a fresh one with a different population distribution).  These three mechanisms should result in observable differences.  For 1, we should see two temperatures even at large cometocentric distances, with very little change in the hotter temperature, and the Phillips band should have more flux than expected because it takes flux from the triplets. For 2, the two temperatures should exist at the optocenter and in the inner coma but should disappear at larger cometocentric distances. Remarkably, the second model is also expected to be independent of the exact parent species \citep{Lambert1990}. Model 3 is somewhat degenerate with 2, but the variations should look different with distance because close to the comet there isn't much ``hot" C$_2$. There has been little to no progress in investigating the bimodal temperature oddity in C$_2$ since these models were proposed. These mechanisms are readily testable by examining spectra at different heliocentric and cometocentric distances, which we will be able to do in future work using the Boltzmann plots and PGOPHER models described in this paper.

The lack of a strong two temperature signal in the tail spectra of both de~Vico and Ikeya-Zhang rules out the first mechanism. It could be argued that mechanism 1 plays some role, but it does not appear to be the root cause of these temperatures. The Boltzmann plot method generalizes nicely to other observations. We intend to further investigate which of the remaining two pathways is important by using Hale-Bopp as a case study because we have high SNR data over a wide range of cometocentric and heliocentric distances. 

The exact values of the temperatures are unclear at best. Knowing the problems with each method, a case can be made for one or the other being correct at some level. If perturbations and calibration errors were accounted for, PGOPHER would be superior but, as it stands, many high J lines, especially in weaker bands, remain poorly fit in the available models. 

There are several limitations to using the Boltzmann plots. First, the rotational temperature is degenerate with vibrational temperature. This can be overcome with enough clean lines of each band, allowing a vibrational temperature to be estimated, assuming similar rotational temperatures for each band. Second, requiring that blended lines have similar excitation energies is incorrect in many cases. This problem is highly dependent on spectral resolution.
Third, as noted above, Boltzmann plots make an assumption that the molecules follow a Boltzmann distribution for the gas.

Using PGOPHER lacks prescriptive power for the population function. Thus, we use the Boltzmann plot to motivate what function we should fit with PGOPHER. This is plausible based on the qualitative agreement with the Boltzmann plots. If the two methods converge, we can use aspects of either when it suits our needs. This would be beneficial for estimating the higher order effects of C$_2$ seen in cometary Swan bands. The error analysis with POGPHER is also largely removed of user influence.

\section{Future Work}
Our knowledge of C$_2$ has continued to improve since \cite{Lambert1990}. An important improvement is the observation of forbidden emission caused by the singlet-triplet intercombinational transitions. \cite{Chen2015} observed some of these emissions, allowing them to calculate the spin orbital coupling of the X$^1\Sigma^+_g$ and b$^3\Sigma^-_g$ states at three vibrational levels. \cite{Gredel1989} demonstrated that these intercombinational transitions are a good candidate for cooling C$_2$ to $\sim 4000$K and possibly causing the two temperatures (see Figure 2 in their paper). \cite{Lambert1990} believed that these transitions are the primary source of uncertainty in the populations. Knowing these values should close the loop on steady state calculations. These steady state spectra can be compared to any observations that have some spatial resolution.

There are ways the models used can be improved. A dedicated effort to determine molecular constants and perturbations so that the PGOPHER models reproduce the higher N levels for all Swan bands would be excellent. It is beyond the scope of the work here, but such a study would remove the guesswork associated with line assignment. Such an investigation might also slightly alter the line strengths used as well. 

One assumption that needed to be made for both the Boltzmann Plot and PGOPHER methods was that the C$_2$ gas had reached equilibrium long before it was observed by us.  \cite{Lambert1983} concluded that the timescale to equilibrium was $\sim$300\,seconds. \cite{Lambert1990} observed the two temperatures were cometocentric-distance dependent and suggested the fluorescence equilibrium model might be insufficient based on that. \cite{Rousselot1994} used a simple Monte Carlo simulation of C$_2$ observed in comet Halley by the Vega spacecraft to study the time evolving low resolution C$_2$ emission. He concluded that $\tau_{\mathrm{eq}}\sim3000$ seconds, an order of magnitude larger than the value used here and elsewhere. This is an intriguing result and, despite enormous advances in computing in the intervening 25 years, has not been explored more. A new attempt to measure this lifetime could be made with observations of 9P/Tempel~1 just after its impact with the Deep Impact spacecraft. In particular, because the (0,0) band should reach equilibrium in about 1/3 of the time for the (1,1) band the temperature evolution might be observable.

To study the influence of an initial population, we could take an approach like \cite{DuckBands} have used, simulating the number of fluorescence cycles needed to reach steady state. There is also a need for more laboratory work like \cite{Jackson1996}. They demonstrated C$_2$ could have two temperatures just from production, but acknowledge that their experimental conditions do not reflect the coma conditions. Perhaps an experiment with more realistic conditions is needed to check the earlier laboratory results. 

In summary, we used high spectral-resolution observations of two comets to show that the cometary C$_2$ gas can be described with bimodal rotational temperatures.  We will employ the techniques described in this paper, along with our substantial high spectral resolution observation database, to explore how factors such as heliocentric distance and cometocentric distance affect what we see.

\acknowledgements
The observations reported in this paper were obtained at The McDonald Observatory, operated by The University of Texas at Austin.  This work was supported by NASA Grant NNX17A186G.

\facility{Smith(Tull 2DCoude spectrograph)}

\software{pwlf \citep{pwlf}, PGOPHER \citep{Colin2017}}

\bibliography{C2Paper}

\listofchanges
\end{document}